\documentclass[12pt,a4paper,twoside,fleqn]{article}
\usepackage{latexsym}
\usepackage{emlines}
\usepackage{epsf}

\begin{document}


\thispagestyle{empty}
\centerline
{\Large Next--to--Leading Order QCD corrections}
\centerline
{\Large for the $B^0 \overline{B^0} $-mixing with an extended Higgs sector }

\vspace{0.5cm}

\centerline
{J. Urban, F. Krauss, U. Jentschura, G. Soff}
\vspace{5mm}

\centerline
{Institut f$\ddot{\rm u}$r Theoretische Physik}
\centerline
{Technische Universit$\ddot{\rm a}$t Dresden}
\centerline
{Mommsenstr.13, D-01062 Dresden, Germany}

\vspace{0.5cm}

\begin{center}
\parbox{15cm}{
{\Large Abstract}

\vspace{0.5cm}

{\footnotesize

We present a calculation of the $B^0 \overline {B^0}$--mixing including 
Next--to--Leading Order (NLO) QCD corrections within the Two Higgs 
Doublet Model (2HDM). The QCD corrections at NLO are contained in the factor 
denoted by $\eta_2$ which modifies the result
obtained at the lowest order of perturbation theory. In the Standard Model 
case, 
we confirm the results for $\eta_2$ obtained by Buras, Jamin and Weisz  
\cite{BJW90}. 
The factor $\eta_2$ is gauge and renormalization prescription 
invariant and it does not depend on the infrared behaviour of the 
theory, which constitutes an important test of the calculations.
The NLO--calculations within the 2HDM enhance the LO--result
up to $18\,\%$, which affects the correlation between $M_H$ and $V_{td}$.}}
\end{center}

\vspace{0.5cm}
\noindent
PACS-numbers: 12.15.Ff; 12.60.Fr; 14.80.Cp; 12.38.Bx \\[0.3cm]
Key--words: Mixing; Oscillations; 2HDM; CP-violation; CKM-matrix 

\section{Introduction}

The $B$-meson plays an important role in present day physics. 
$B^0 \overline{B^0} $--mixing as well as 
$B$-meson decays can be used to determine the CKM--elements and 
to investigate $CP$-violation within the Standard Model (SM).
Two large experiments, namely the BABAR-Collaboration at SLAC \cite{BAB95} 
and the Belle Collaboration at KEK \cite{BEL94}, will start taking data
in the near future. The major subject of research in those experiments 
is the determination of the CKM--matrix elements. This may also 
provide first hints on physics beyond the Standard Model.

\begin{figure}[h]
\begin{tabular}{cc}
\centerline{\mbox{\epsfxsize=6cm\epsffile{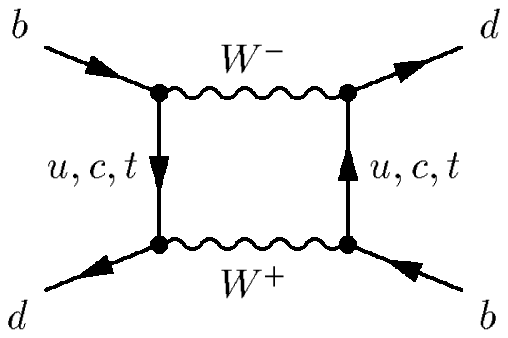}}}
\hspace{-4cm}{\mbox{\epsfxsize=6cm\epsffile{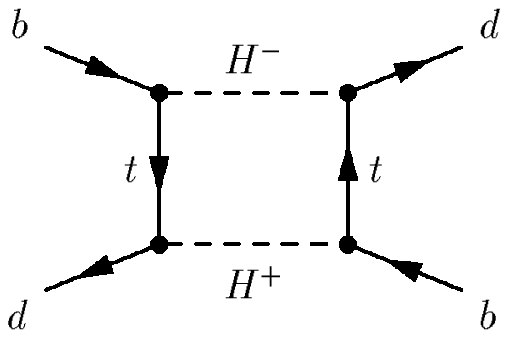}}}&
\end{tabular}
\centerline{\mbox{\epsfxsize=6cm\epsffile{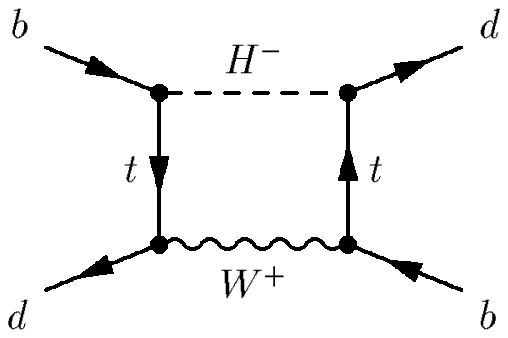}}}
\caption{\footnotesize
Box diagrams for the $B^0\overline {B^0}$--mixing in the framework of
2HDM.  We have also taken  into account crossed  
diagrams, which are related to the original ones by a Fierz--transformation.
\label{LOboxen}}
\vspace{1cm}
\begin{tabular}{cc}
\centerline{\mbox{\epsfxsize=6cm\epsffile{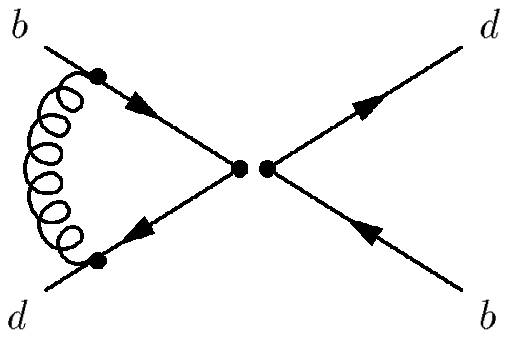}}}&
\hspace{-4cm}{\mbox{\epsfxsize=6cm\epsffile{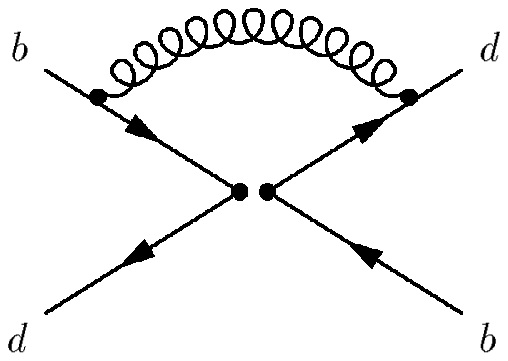}}}\\
\centerline{\mbox{\epsfxsize=6cm\epsffile{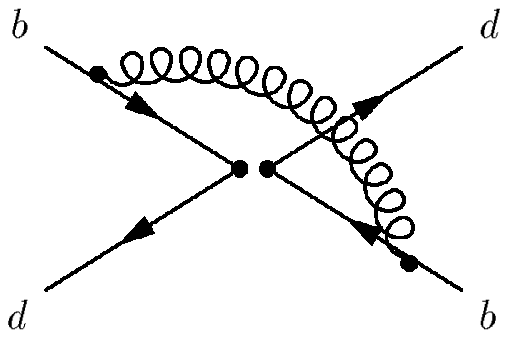}}}&
\hspace{-4cm}{\mbox{\epsfxsize=6cm\epsffile{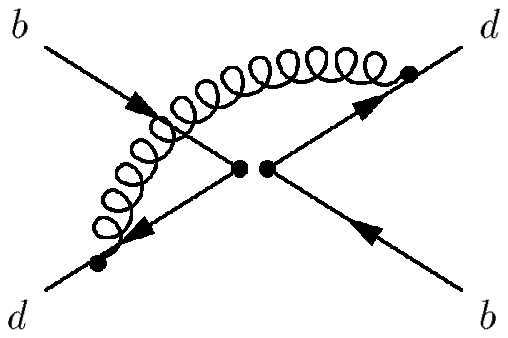}}}
\end{tabular}
\caption{\footnotesize
QCD-corrections to the effective four quark interaction.
They are needed
in the calculation of the anomalous dimension at LO and for the correct
separation of long and short distance contributions.\label{effekt}}
\end{figure}

In general terms $B^0 \overline{B^0} $--mixing is a Flavour Changing 
Neutral Current (FCNC) process generated through weak interactions. 
In the SM, the process is generated at the lowest order of perturbation
theory via the box diagrams displayed in 
Fig. \ref{LOboxen}. Due to the large difference in the masses of 
heavy particles in the box ($W^\pm$, $H^\pm$ and $t$--quarks) 
and much lighter external particles ($b$ and $d$--quarks), 
it is possible to disentangle long and short distance effects. 
This can be accomplished within an effective theory.
The proper separation of short and long distance QCD-effects at
Next--to--Leading Order was first presented for the SM in ~\cite{BJW90}. 
We are utilizing here the same framework of renormalization group improved 
perturbation theory.

Our purpose is to include two Higgs--boson doublets as an extended 
Higgs--boson sector. Such models are usually called 
Two Higgs Doublet Models (2HDM) ~\cite{{HIG90},{BURD95}}. 
Two particular models which are usually considered (so--called Model I 
and Model II) differ in the couplings of the charged Higgs--bosons 
to fermions (see for example \cite{HIG90}):
\begin{eqnarray}
{\cal L}_{H\,ud}^{I} &=& \frac{g_W V_{ud}}{\sqrt{2}m_W}\left[
m_u \cot(\beta) P_L - m_d \cot(\beta) P_R\right]\;,\\
{\cal L}_{H\,ud}^{II} &=& \frac{g_W V_{ud}}{\sqrt{2}m_W}\left[
m_u \cot(\beta) P_L + m_d \tan(\beta) P_R\right]\;.
\end{eqnarray}
Here $g_W$ is the weak coupling constant, $m_W$ is the $W$--mass,
and the projectors $P_{L,R}$ are defined by
\begin{equation}\label{Projs}
P_{L,R}=\frac{1\mp \gamma_5}{2}\;.
\end{equation}
The subscripts $u$ and $d$ denote up-- and down--type
quarks, respectively. 

As we neglect all quark masses except $m_t$, both models are identical in
our case. Thus our calculations are valid for all values of
$\tan\beta$ for the Model I, but only for $\tan\beta<<m_t/m_b$
in the Model II.
Looking at the Feynman rules for the Higgs--quark--quark 
coupling, it is easy to verify that in Model II, the term 
proportional to $m_b$ becomes important for large values of $\tan\beta$. 
Due to the
projectors in the vertex, we would end up with different operators
for the $\tan\beta$ and $\cot\beta$ parts. However, it should be stressed
here that Model II is more interesting, because this is
the choice favoured by the Minimal Supersymmetric Standard Model
\cite{HIG90,HAB9x}. Large values of $\tan\beta$ will be considered in a 
future publication.

We continue our discussion with a short review of the Leading Order (LO)
calculations. 
The basic idea is rather simple. First, one has to calculate the 
box diagrams in Fig.\ref{LOboxen}. 
They can be evaluated in the Feynman--'t Hooft gauge for the
$W$--boson. This leaves us with a physical $W$--boson and a 
would-be Goldstone boson (unphysical scalar Higgs boson)  
of the same mass and with couplings proportional to the quark masses.
As all quark masses except the top--quark mass are equal to zero to
a good approximation, $t$ is the only active flavour in 
boxes containing Higgs particles or unphysical scalars. 
In the remaining box diagrams, the effects of u- and c- quarks are taken into 
account by the GIM mechanism. 
Evaluating the box diagrams in this framework leads to the well 
known Inami--Lim functions ~\cite{IL81}--\cite{GLA87}.

The scaling behaviour from the matching scale $\mu_0=m_W$ of the full and 
effective theory down to a lower scale 
is then determined by the one--loop QCD correction to the effective vertex
generated by the previous procedure of integrating out the internal
heavy degrees of freedom. The renormalization group equation gives us the 
factor $\overline\eta_{LO}$ describing the effect of scaling
~\cite{GW80}.

The effective Hamiltonian for $B^0 \overline{B^0} $--mixing at LO
reads   
\begin{equation}
\label{heff}
H_{\rm{eff}} \;= \frac14\,{G^2_F \over \pi^2}\; m_W^2 \;{( V_{td}V^\ast_{tb})}^2 
\;\overline\eta_{LO}\; S(x_W,x_H) \;\hat{\cal O}_{LL}\;,
\end{equation}
where
\begin{equation}
\hat{\cal O}_{LL} = \biggl[\;\bar{d} \;\gamma_{\mu} \;
P_L\; b \;\biggr] \;\biggl[\; \bar{d}\; \gamma ^{\mu}\; 
P_L\;b\; \biggr]\;
\end{equation}
and
\begin{equation}
S(x_W,x_H)\;=\;S_{WW}(x_W)+2\,S_{WH}(x_W,x_H)+S_{HH}(x_H)\;.
\end{equation} 
The LO Inami--Lim functions $S_{WW}(x_W)$, $S_{WH}(x_W,x_H)$ and $S_{HH}(x_H)$ 
can be found in Appendix A. The arguments of the 
Inami--Lim--functions will be denoted $x_{W,H}=m_t^2/m_{W,H}^2$. 

The factor $\overline\eta_{LO}$ is determined by the $\beta$--function 
\cite{FIE89} and the 
anomalous dimension $\gamma$ to be obtained by 
evaluating the diagrams of Fig. \ref{effekt}. It reads
\begin{equation}\label{eta0}
\overline\eta_{LO}\;=\;
\biggl[ \;{{\alpha_S(m_W)}\over {\alpha_S(m_B)}}\;
\biggr] ^{\gamma^{(0)}/(2 \beta_0)} \;=\;
\biggl[ \;{{\alpha_S(m_W)}\over {\alpha_S(m_B)}}\;
\biggr] ^{6/23}\approx0.85\; .
\end{equation}
For later convenience we define $\eta_{LO}=\alpha_S(m_W)^{(6/23)}$.

To evaluate the matrix element of $H_{\rm{eff}}$ we have to employ
\begin{equation}\label{operator}
\langle\bar{B}^0| \hat{\cal O}_{LL}| B^0 \rangle\; 
=\; {2 \over 3} \;B_B \;(\mu) \;f^2_B \;m^2_B \;,
\end{equation}
where $f_B$ is the $B$--meson decay constant and $B_B(\mu)$ parameterizes 
deviations from the vacuum insertion 
approximation.

\section{Explicit QCD-Corrections}

\mathindent0pt

\begin{figure}[h]
\begin{center}
\begin{tabular}{cc}
\mbox{{\epsfxsize=6cm\epsffile{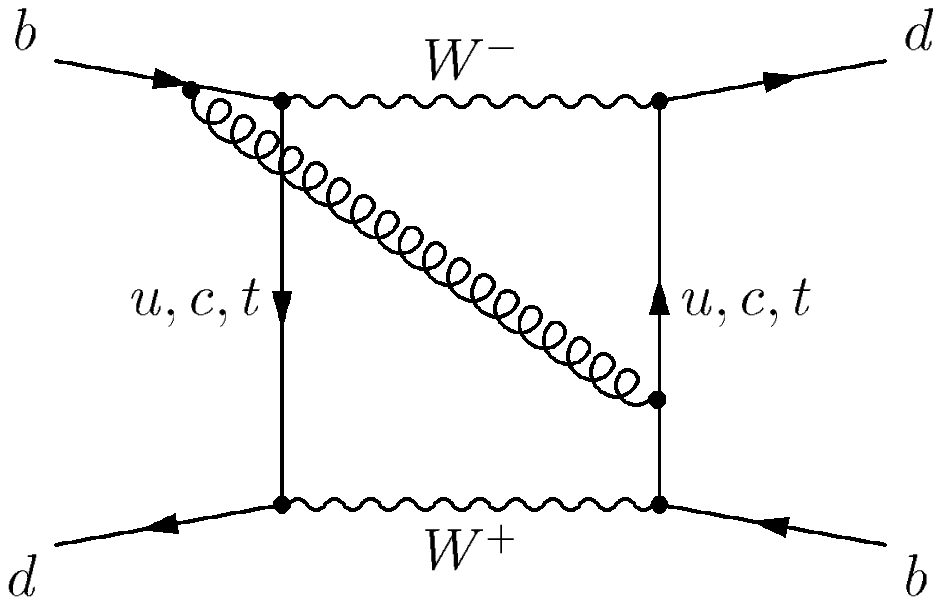}}}
&
\mbox{{\epsfxsize=6cm\epsffile{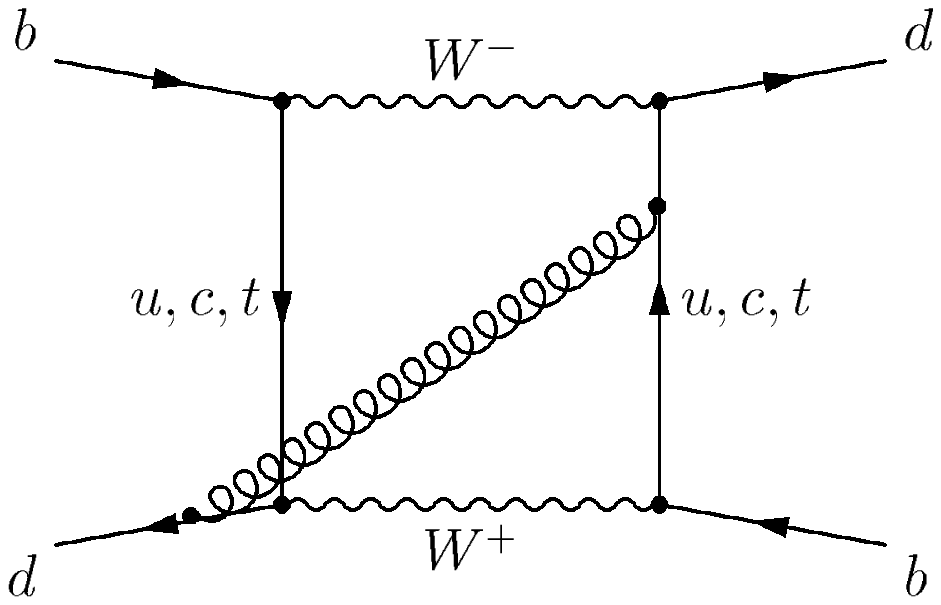}}}
\\
a)&b)\\[1cm]
\mbox{{\epsfxsize=6cm\epsffile{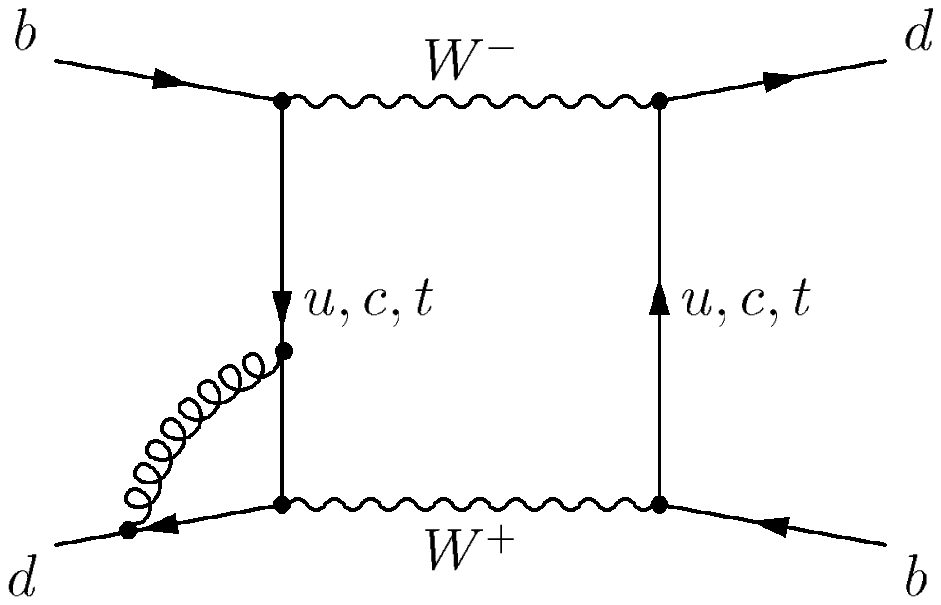}}}
&
\mbox{{\epsfxsize=6cm\epsffile{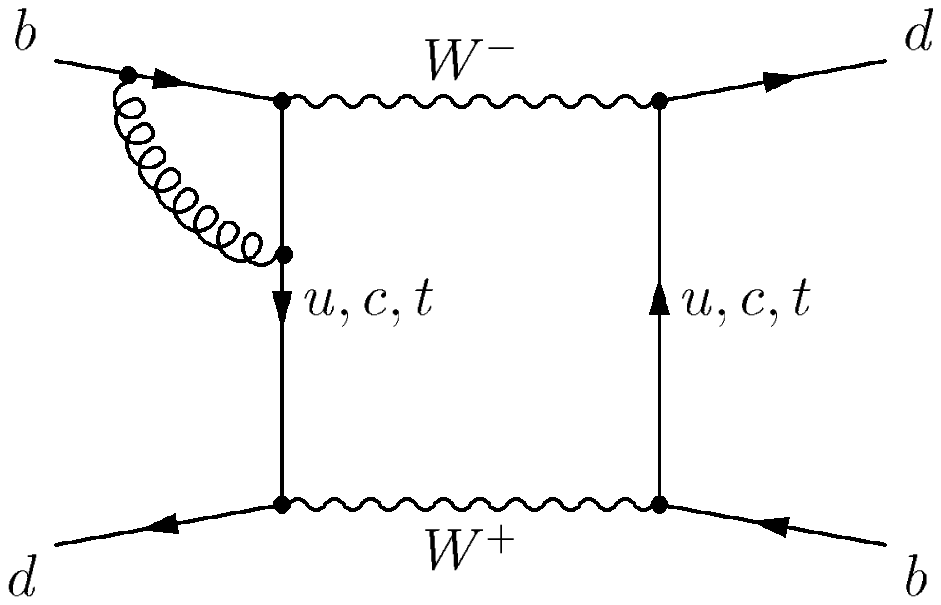}}}
\\
c)&d)\\[1cm]
\mbox{{\epsfxsize=6cm\epsffile{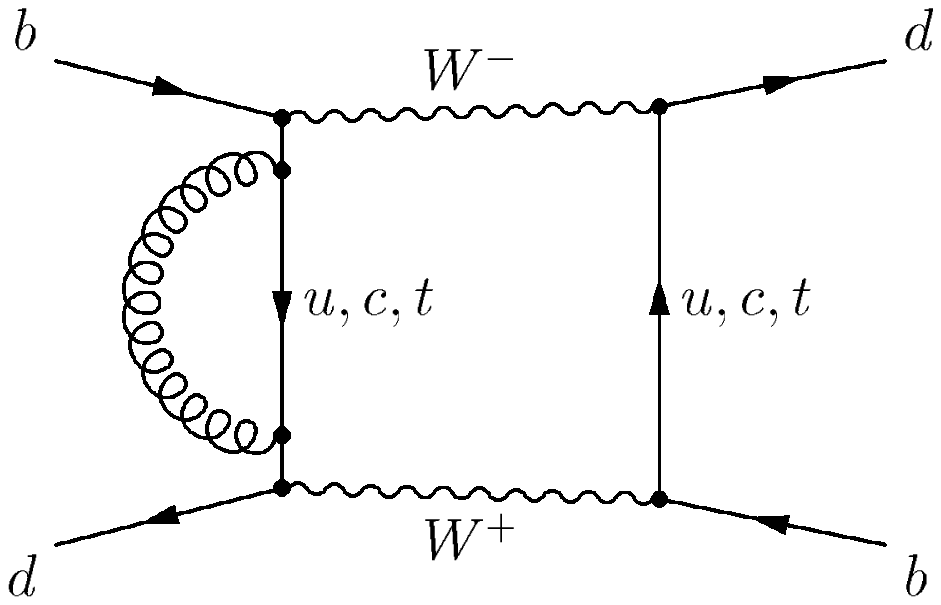}}}
&
\mbox{{\epsfxsize=6cm\epsffile{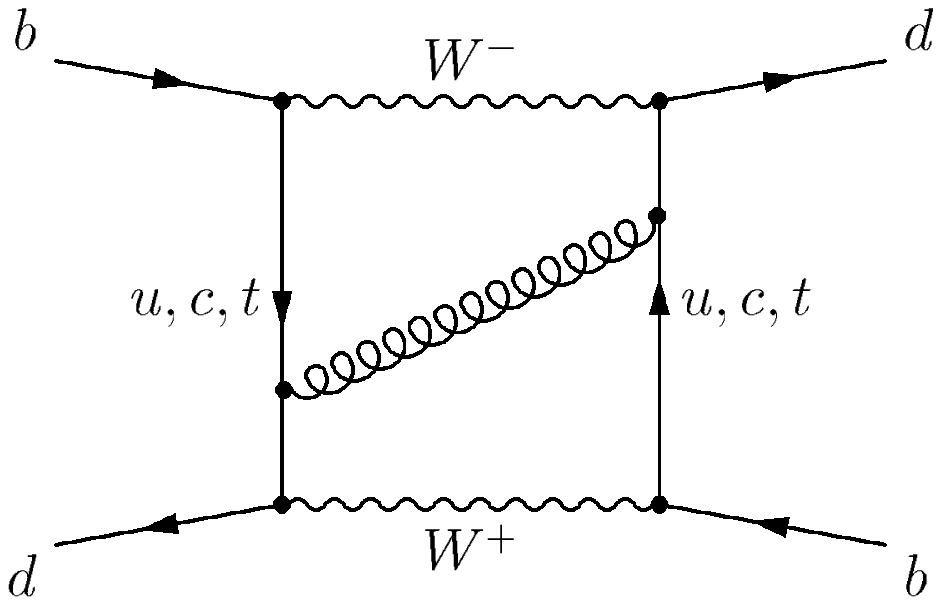}}}
\\
e)&f)
\end{tabular}
\end{center}
\caption{\footnotesize
Diagrams for the NLO-calculation in the SM, which are convergent
in the limit of vanishing external masses. Additional diagrams
with one Higgs-boson and two Higgs-bosons have to be calculated
in the 2HDM. Furthermore, we have to consider crossed diagrams.
\label{NLOboxen1}}
\end{figure}

\begin{figure}[h]
\begin{center}
\begin{tabular}{cc}
\mbox{{\epsfxsize=6cm\epsffile{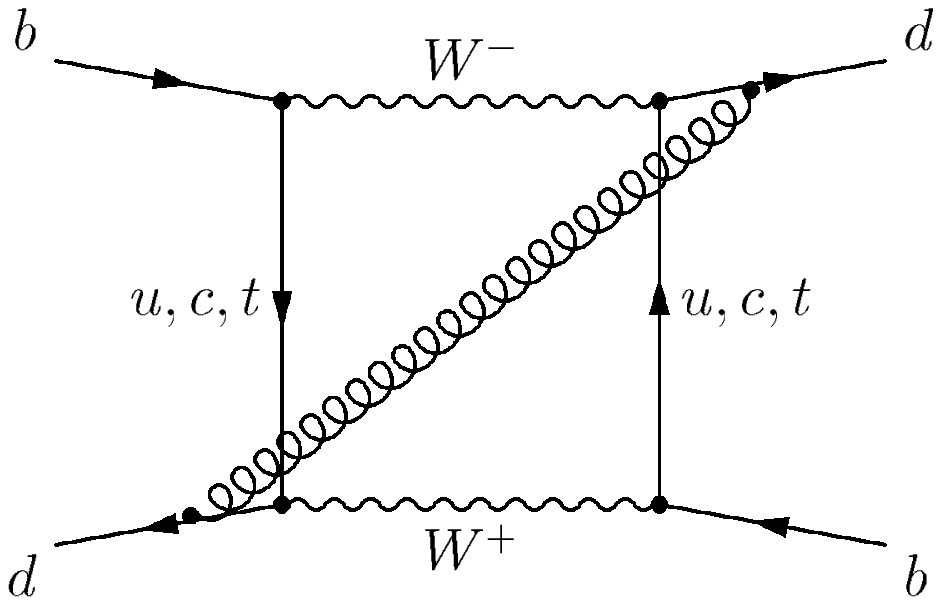}}}
&
\mbox{{\epsfxsize=6cm\epsffile{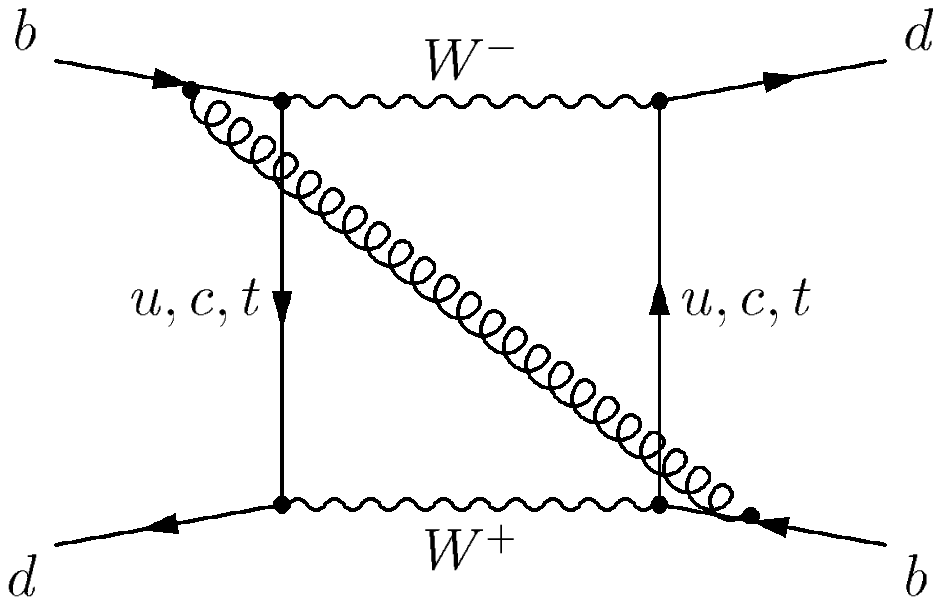}}}
\\
g)&h)\\[1cm]
\mbox{{\epsfxsize=6cm\epsffile{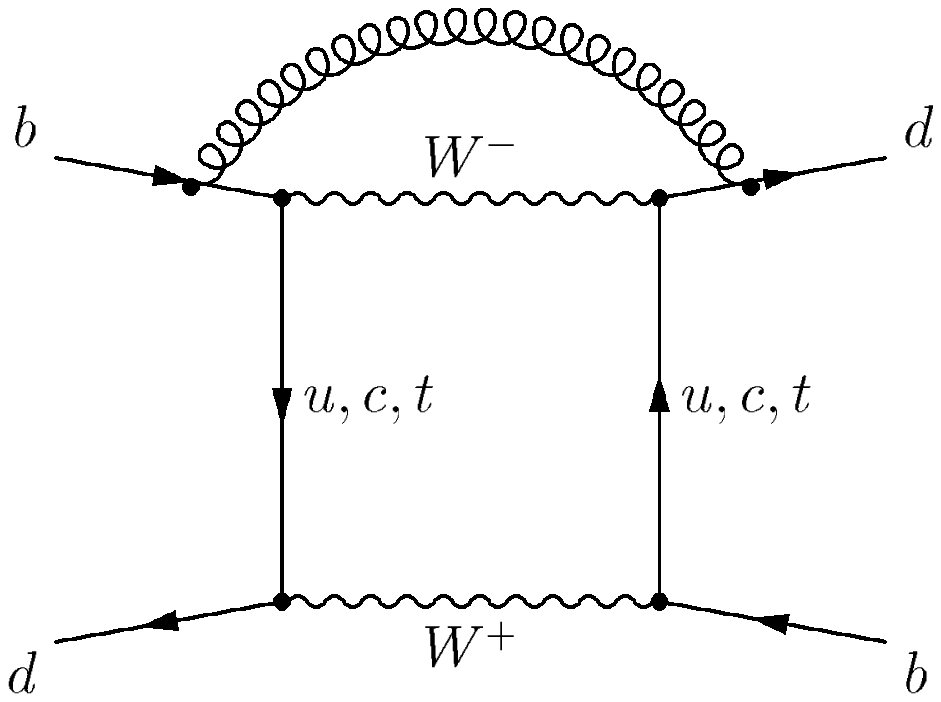}}}
&
\mbox{{\epsfxsize=6cm\epsffile{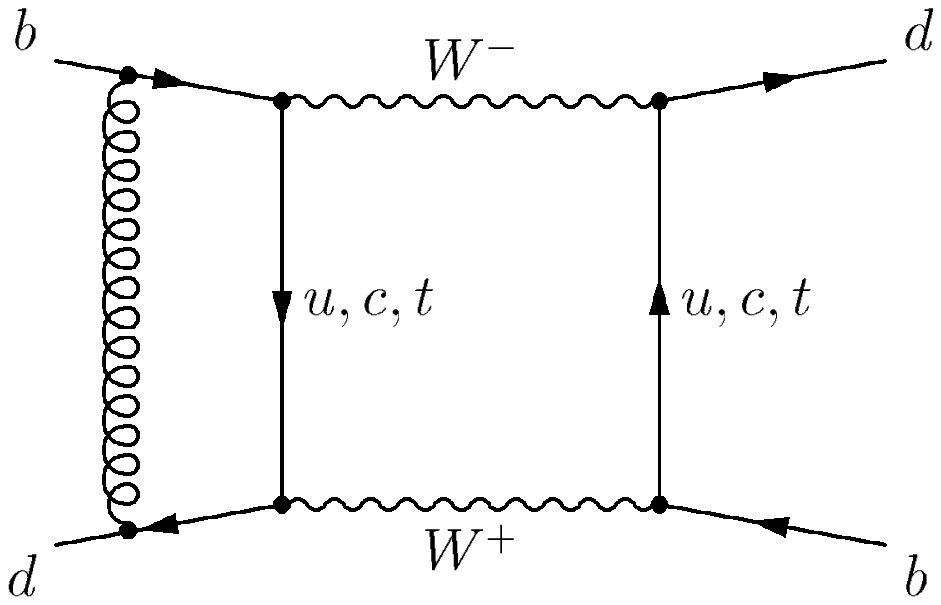}}}
\\
i)&j)\\[1cm]
\mbox{{\epsfxsize=6cm\epsffile{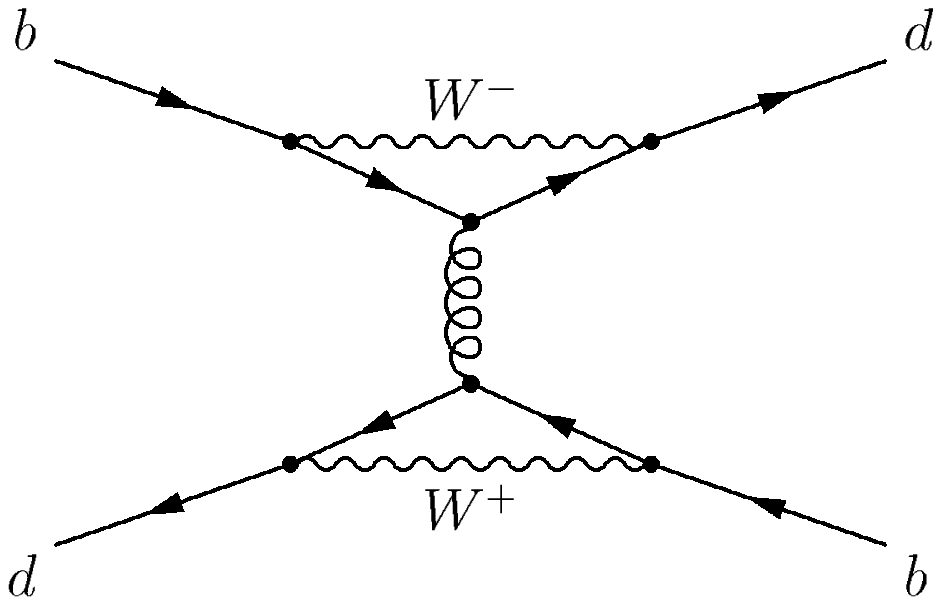}}}
&
\\
k)&
\end{tabular}
\end{center}
\caption{\footnotesize
Diagrams for the NLO-calculation in the SM, in which
infrared divergences appear when one sets
external masses to zero. The last diagram is 
the so-called double-penguin, which does not contribute
to the mixing at the leading order in $(m_B/m_W)^2$.
Additional diagrams
with one Higgs-boson and two Higgs-bosons have to be calculated
in the 2HDM. Furthermore, we have to consider crossed diagrams.
\label{NLOboxen2}}
\end{figure}

The perturbative result for ${\cal O}(\alpha_S)$ corrections in 2HDM
will be presented in this section. The results can be obtained by 
evaluation of the diagrams of Figs. \ref{NLOboxen1} and \ref{NLOboxen2}.
We performed our calculation in an arbitrary covariant $\xi$-gauge 
for the gluon and
we employed the Feynman--'t Hooft gauge for the $W$--boson.

As one easily notices, the diagrams a, b and f - i have the "´octet"--structure 
$\hat T_a \otimes \hat T^a$, whereas the diagrams c - e and j have 
"´singlet"--structure $\hat{\bf 1} \otimes \hat{\bf 1}$ 
in colour space. The double penguin diagram k contributes
to the considered process at the order of ${\cal O} (1/m_W^4)$ which is
negligible for our purpose.

Diagrams containing vertex-- and self--energy corrections (diagrams 
c, d and e) lead to UV-divergent integrals and hence they have to be 
regularized and renormalized. We are using dimensional regularization
with anticommuting $\gamma_5$. This corresponds to the NDR--scheme
(``naive'' dimensional regularization scheme) 
\cite{{BUR90a},{HER}}. We performed the necessary renormalization in
the $\overline{MS}$--scheme \cite{{BUR90a},{BAR78}}. All other 
diagrams are UV--finite.
Furthermore, one notices that the diagrams g - j contain 
infrared divergences. To deal with them we keep the external 
quark masses whenever necessary. As we will see, the external quark masses 
do not affect the final result for the
Wilson--coefficient. This justifies the method we have chosen.

We could also treat the infrared divergences with the method of dimensional 
regularization ~\cite{COL}, but our intention was to verify
the results provided by Buras and collaborators in ~\cite{BJW90} and
to extend these calculations to the 2HDM. 

The ${\cal O}(\alpha_S)$-corrections to the Hamiltonian (\ref{heff})
have the following structure
\begin{equation}
\label{deltaheff}
\langle\Delta H_{\rm{eff}}\rangle \;= \frac14\,{G^2_F 
\over \pi^2}\; m_W^2 \; { \alpha_S \over 
4 \pi}\;  (V_{td}V^\ast_{tb})^2 \;  U(x_W,x_H)\;,
\end{equation}
where 
\begin{equation}
\label{usm}
U(x_W,x_H) = \sum_k \; \biggl( C_F\, \hat{\bf 1} 
\otimes \hat{\bf 1}\, \phi^{(1)}_k (x_W,x_H) + \hat T_a \otimes \hat T^a\, 
\phi^{(8)}_k (x_W,x_H) \biggr) \langle\hat{\cal O}_k\rangle
\end{equation}
with $k\in \{ LL,1,2,3 \}$.  
$C_F$ is the colour-factor defined by $C_F=(N_c^2-1)/(2N_c)$ and $N_c$
is the number of colours. 

The operators $\hat{\cal O}_i$ read

\begin{eqnarray}
\hat{\cal O}_{LL} &=& \left[\bar{d} \;\gamma_{\mu}\,P_L \;b\right]
                    \left[\bar d \;\gamma^{\mu}\,P_L \;b\right]\;,
\\[2mm]
\hat{\cal O}_1 &=& \left[\bar{d} \;P_L\; b\right] \left[\bar d
                 \;P_L\; b\right] - {1\over 4}\left[\bar{d} 
                 \;\sigma_{\mu \nu}\,P_L\; b\right]
                 \left[\bar d\;\sigma^{\mu \nu}\,P_L \;b\right]+ 
                 (P_L \longrightarrow P_R)\;, \;\\[2mm]
\hat{\cal O}_2 &=& \left[\bar{d} \;P_L\; b\right]  
                 \left[\bar d \;
                 P_R\; b\right] +\left[\bar{d} \;P_R\; b\right]
                 \left[\bar d  \;P_L\; b\right]\;,\\[2mm]
\hat{\cal O}_3 &=& \left[\bar{d} \;\gamma_{\mu}\,P_L\; b\right] 
                 \left[\bar d \;\gamma^{\mu}\,P_R \; b\right]+ 
                 ( P_L\longleftrightarrow P_R )\;.
\end{eqnarray}

The operator $\hat{\cal O}_{1}$ steems from diagrams g and h in 
Fig. \ref{NLOboxen2}, 
whereas the operators  $\hat{\cal O}_{2}$ and $\hat{\cal O}_{3}$ follow from
diagrams i and j, respectively.
Since the relevant operator $\hat{\cal O}_{LL}$ is self--conjugate 
under Fierz--transformations, we will
use
\begin{equation}\label{singoct}
\hat T_a \otimes \hat T^a = C_A \hat{\bf 1} \otimes \hat{\bf 1} 
= \frac12 \left(1-\frac{1}{N_c}\right) \hat{\bf 1} \otimes \hat{\bf 1} 
\;
\end{equation}
to retain only one operator. Nevertheless, to keep the calculations
transparent, we will abandon the distinction between octet and singlet 
at the very end only.

The coefficient functions $\phi$ in equation (\ref{usm}) can be
decomposed as
\begin{equation}
\noindent\phi^{(i)}_j = \chi^{(i,SM)}_j (x_W) + \chi^{(i,H)}_j (x_W,x_H)\;. 
\end{equation}
The functions $\chi^{(i,SM)}_j (x_W)$ were already given in ref. ~\cite{BJW90}. 
We have recalculated them and we confirm these Standard Model results. 
\begin{eqnarray}
\noindent\label{chi1}\nonumber
\chi^{(1,SM)}_{LL}(x_W)&=& L^{(1,SM)}(x_W) + \left[2\,\xi\ + 2\,\xi\,g_{IR}\, 
+ 2\,\xi\,\ln (x_{\mu_0}) + 6\,\ln (x_{\mu_0})\; x_W \,{\partial \over \partial x_W} 
\right] S_{WW} (x_W) \;,\\ \\
\chi^{(8,SM)}_{LL}(x_W)&=& L^{(8,SM)}(x_W) + \left[2\, \xi\, + (3+\xi)
 \ln (x_b \,x_d) \, + 2\,\xi\, g_{IR}\right] S_{WW} (x_W) \;,\label{chi1a}\\
\chi^{(8,SM)}_1(x_W)&=& - (3+\xi)\, S_{WW}(x_W) \;,\label{chi1b}\\
\chi^{(8,SM)}_2(x_W)&=& - 2\, \chi^{(1,SM)}_3(x_W)\; =
-(3+\xi) {m_b \,m_d \over {m_d^2 - m_b^2}} 
\ln \left( x_d \over x_b\right) \, S_{WW}(x_W)\;,\label{chi1c}
\end{eqnarray}
with all other $\chi$ equal to zero.

We have introduced the following abbreviations:  
\begin{eqnarray}
x_{W,H} &=& \frac{m_t^2}{m_{W,H}^2} \;,\;\;\; 
x_{d,b}  = \frac{m_{d,b}^2}{m_W^2} \;,\;\;\; 
x_{\mu_0} = \frac{\mu_0^2}{m_W^2}\;,\\ 
g_{IR} &=& - \, { x_d \ln (x_d) - x_b \ln (x_b) \over x_d-x_b } \;.
\end{eqnarray}
Here, $m_t$ stands for the top quark mass renormalized at the scale 
$\mu_0$ in the $\overline{MS}$-scheme.

The function $\chi^{(1,H)}_{LL}$ reads

\begin{eqnarray}
\noindent\label{chi}\nonumber
\chi^{(1,H)}_{LL}(x_W,x_H)&=& L^{(1,H)}(x_W,x_H)\\
\nonumber
& & + \left[2\,\xi\ + 2\,\xi\,g_{IR}\, + 2\,\xi\,\ln (x_{\mu_0})
            + 6 \,\ln (x_{\mu_0}) \sum_{i=H,W} x_i 
            \frac{\partial}{\partial x_i}\right]\\
& & \qquad\qquad \cdot (2\;S_{WH} (x_W,x_H)\;+\;S_{HH}(x_W,x_H))\,.
\end{eqnarray}
The analytical expressions for the functions $S_{WW}, S_{WH} , S_{HH}$ and $L^{(i,SM,H)}$  
are listed in Appendix A. Note that powers of $\tan\beta$ enter
in the definition of the various $S$ and $L$ indicating the
number of Higgs--bosons involved.

The remaining newly calculated functions $\chi_{LL}^{(8,H)}$, $\chi_{1}^{(8,H)}$,
$\chi_{2}^{(8,H)}$ and $\chi_{3}^{(1,H)}$ can be obtained respectively from 
$\chi_{LL}^{(8,SM)}\:$ (\ref{chi1a}), $\chi_{1}^{(8,SM)}\:$ (\ref{chi1b}), 
$\chi_{2}^{(8,SM)}\:$ (\ref{chi1c}) and $\chi_{3}^{(1,SM)}\:$ (\ref{chi1c})
by changing $L^{(i,SM)}(x_W)$ to $L^{(i,H)}(x_W,x_H)$ and $S_{WW}(x_W)$ to 
$(2\,S_{WH}(x_W,x_H)+S_{HH}(x_H))$.

We have obtained the function $L^{(i,SM,H)}$ by using computer algebra
systems, especially the program FORM
to evaluate the Dirac--structure and
Mathematica 3.0 ~\cite{mathem} for summarizing all terms. 
Maple ~\cite{maple} was used for some
integrations.

Note that the results of the particular diagrams 
exhibit non--trivial dependence on the gauge--parameter $\xi$ and
the IR--masses, but the functions $L^{(i,SM,H)}$ are gauge 
as well as renormalization prescription invariant. They do not 
depend on the external quark masses either. Terms depending on
$\xi, x_b, x_d$, or $x_{\mu_0}$ are proportional to the LO-Inami-Lim-functions 
$S_{WW}, S_{WH}$ or $S_{HH}$. This reflects the exact 
factorization of long and short distance effects.

\section{Matching and running}
We start this section with the observation that Eq. (\ref{usm})
is obviously unphysical because the coefficient functions $\chi^{(i)}_k$ 
are gauge dependent. Actually this is nothing 
but an artifact of the specific way we have regularized the IR--divergences. 
Other regularization schemes being not based on the use of small 
masses for the 
external quarks would change the obtained result. Dimensional
regularization or a gluon mass, for example, would
leave us with only one operator $\hat{\cal O}_{LL}$ at this stage.
To take proper account of this fact, we have to evaluate the
matrix element of the
physical operator $\hat{\cal O}_{LL}$ up to order ${\cal O}(\alpha_S)$
using the same IR--regularization prescriptions
as before. Yet the one--loop amplitude of $\hat{\cal O}_{LL}$  
as given by Fig. \ref{effekt} results in the same unphysical operators 
$\hat{\cal O}_{1,2,3}$ with the same coefficients. 
\begin{eqnarray}\nonumber\label{Oper1}
\langle\hat{\cal O}_{LL}(\mu_0)\rangle_{\rm 1loop} =
  \langle\hat{\cal O}_{LL}\rangle_{\rm tree}+
  \frac{\alpha_S(\mu_0)}{4\pi}\;
  \sum_k\left[C_F \chi^{(1)}_{\Delta,k}(\mu_0)+
              C_A \chi^{(8)}_{\Delta,k}(\mu_0)\right] 
  \hat{\bf 1} \otimes \hat{\bf 1}
  \langle\hat{\cal O}_{k}\rangle_ {\rm tree}\;,\\
\end{eqnarray}
where the sum over $k=LL,1,2,3$ is understood. The functions 
$\chi^{(i)}_{\Delta,k}(\mu_0) $ are:

\begin{eqnarray}
\chi^{(1)}_{\Delta,LL}&=&-3+2\xi\ln(x_{\mu_0})+2\xi-2\xi\frac1{x_d-x_b}
                             \left[x_d\ln(x_d)-x_b\ln(x_b)\right]\;,\\
\chi^{(8)}_{\Delta,LL}&=&-6\ln(x_{\mu_0})-5+2\xi+
                              (3+\xi)\ln(x_d x_b)-2\xi\frac1{x_d-x_b}
                             \left[x_d\ln(x_d)-x_b\ln(x_b)\right]\;,\\
\chi^{(1)}_{\Delta,3}&=&-\frac12 \chi^{(8)}_{\Delta,2}=
                             \frac{(3+\xi)}{2}\frac{m_d m_b}{m_d^2-m_b^2}
                             \ln\left(\frac{x_d}{x_b}\right)\;,\\
\chi^{(8)}_{\Delta,1}&=&-(3+\xi)\;.
\end{eqnarray}

Eq. (\ref{Oper1}) allows us to write for the matrix element of the 
effective Hamiltonian up to order ${\cal O}(\alpha_S)$ in the following form:
\begin{eqnarray}\nonumber
\langle H_{\rm{eff}}\rangle & =&  \langle H_{\rm{eff}}  ^{(0)} 
                        +  \Delta H_{\rm{eff}} ^{(1)}\rangle \\ 
              & =& \frac14\,{G^2_F \over \pi^2}\; m_W^2 \;  
                    {( V_{td}V^\ast_{tb})}^2 \; 
                    C_{LL} (\mu_0) 
                    \, \langle\hat{\cal O}_{LL}(\mu_0)\rangle_{\rm 1loop}
\label{matrixelem}
\end{eqnarray}
with $\langle O_{LL}(\mu_0)\rangle_{\rm 1loop}$ as given 
in Eq. (\ref{Oper1}) and $C_{LL} (\mu_0)$ reads:
\begin{equation}
C_{LL} (\mu_0) = S(x_W,x_H)+{\alpha_S(\mu_0) \over 4 \pi} 
\;D(x_W,x_H,x_{\mu_0}) \;.\label{coeff}
\end{equation}
The change due to the inclusion of the charged physical Higgs
results in replacing $S_{WW}(x_W)$ and $D_{SM}(x_W)$ by
\begin{equation}
S_{\rm 2HDM}(x_W,x_H)= S_{WW}(x_W) +
                    2\, S_{WH}(x_W,x_H) + S_{HH}(x_H)\;\;,
\end{equation}
\begin{equation}
D_{2HDM}(x_W,x_H,x_{\mu_0}) = D_{SM}(x_W,x_{\mu_0}) +
                     D_{H}(x_W,x_H,x_{\mu_0})\;\;,
\end{equation}
where
\begin{eqnarray}\nonumber\label{D1}
D_{SM}(x_W,x_{\mu_0}) &=& C_F \left\{ L^{(1,SM)}(x_W) 
+ \biggl[ 6 \ln (x_{\mu_0})  
\; \biggl( x_W {\partial \over \partial x_W} \biggr) 
+3 \biggr] \; S_{WW}(x_W) \right\}\\ 
& & + C_A  \left\{ L^{(8,SM)}(x_W) + 
\biggl[ 6 \ln (x_{\mu_0})  +5 \biggr] \; S_{WW}(x_W) \right\}\;\;,
\end{eqnarray}
\newpage
\begin{eqnarray}
\nonumber
D_{H}(x_H,x_W,x_{\mu_0}) &=& C_F \left\{ L^{(1,H)}(x_W,x_H)\right.\\
& &\nonumber
\left.+\left[6 \ln (x_{\mu_0}) \sum_{i=H,W} x_i {\partial \over \partial x_i}
+3\right] \biggl(
2\; S_{WH}(x_W,x_H) + S_{HH}(x_W,x_H)\biggr)\right\}\\
& &\nonumber
+ \;C_A \left\{ L^{(8,H)}(x_W,x_H) \right.\\
& &
+ \left.\biggl[ 6 \ln (x_{\mu_0})  +5 \biggr] 
\bigl(2\, S_{WH}(x_W,x_H) + S_{HH}(x_W,x_H) \bigr) \right\}\,. \label{D2}
\end{eqnarray}
To obtain this, we have utilized Eq. (\ref{singoct}) and the 
behaviour of $\hat{\cal O}_{LL}$ under Fierz--transformation.
The finite quantities in front of the LO Inami--Lim functions 
in Eqs. (\ref{D1}) and (\ref{D2}) are the
remnants of the functions $\chi^{(i)}_{\Delta,k}(\mu_0)$ of Eq. (\ref{Oper1}).
The coefficient $C_{LL}(\mu_0)$ in Eq. (\ref{matrixelem}) exhibits 
no dependence on the choice of the gauge and the external quark states. 

What remains to be done is the running of our effective theory
down to a lower scale. In \cite{BJW90} it was shown
that a good choice for this matching scale is $\mu_0 = m_W$ resulting
in $x_{\mu_0} = 1$ and that the physical
observables do not depend on the choice of this scale.  
The scaling down from this matching scale to the mesonic scale is pursued
by means of the renormalization group equation.  
The renormalization group equation for the Wilson coefficient
$C_{LL}(\mu)$ in the SM reads
\begin{equation}
\label{rge1}
\left[\mu\frac{d}{d \mu}-\gamma(g)\right]
C_{LL}(\mu)=0\;\;,
\end{equation}
with the initial condition $C_{LL}(\mu_0=m_W)$ given in Eq. (\ref{coeff}).

Expansion of the anomalous dimensions and the
$\beta$--function yields
\begin{eqnarray}
\label{beta}
\beta(g)&=&-\beta_0\,\frac{g^3}{(4\pi)^2}-\beta_1\,\frac{g^5}{(4\pi)^4}-
             \ldots\;\;,\\
\gamma(g)&=&\gamma^{(0)}\,\frac{g^2}{(4\pi)^2}+
            \gamma^{(1)}\,\frac{g^4}{(4\pi)^4}+\ldots\;\;.
\label{anomdim}
\end{eqnarray}
The coefficient $\gamma^{(1)}$ was calculated by Buras and Weisz 
in \cite{BUR90a}. Note that $\gamma^{(1)}$ is renormalization scheme dependent 
\cite{BUR90a,ALT81a,ALT81b}. The coefficients relevant in our further 
calculation are
\begin{eqnarray}
\label{3eck}
\beta_0 &=& {1 \over 3} \; (11\; N_c - 2 \;n_f) \;\;,\\  
\quad \beta_1 &=& {34 \over 3} \; N_c^2 - {10 \over 3} \;
N_c\; n_f - 2 \; C_F \; n_f \;\;,\\
\gamma^{(0)} &=& 6 \; { N_c - 1 \over  N_c}  \;\;,\\
\gamma^{(1)} &=& {N_c-1 \over 2 \;N_c} \; \biggl[
-21 + {57 \over N_c} - {19 \over 3} \; N_c + {4 \over 3} \;n_f \biggr]
\;\;
\end{eqnarray}
where $N_c$ is the number of colours and $n_f$ is the number of active
flavours.

The solution to the renormalization group equation for the Wilson--coefficient
reads 
\begin{equation}
\label{4eck}
C_{LL} (\mu) = {\rm exp}\; \Biggl[-\int^{\bar{g} (m_W)}_{\bar{g} 
(\mu)} \; dg' \; 
{ \gamma(g') \over \beta(g')} \; \Biggr] \; C_{LL} (m_W)\;.
\end{equation}
The evolution from the initial or matching scale $m_W$ to a lower one 
may be performed as in a massless theory because the top quark has 
been integrated out previously and appears in the matching 
condition
only. At the NLO, Eq. (\ref{4eck}) can be approximated by
\begin{eqnarray}
\label{2eck}
C_{LL} (\mu) \approx&& {\rm exp}\; \Biggl[\int^{\bar{g} 
(m_W)}_{\bar{g} 
(\mu)} \; dg'
\; {\gamma^{(0)}\; { g^2/(4 \pi)^2 } + \gamma^{(1)}\; { g^4/(4 \pi)^4 }  
\over \beta_0 \; { g^3/(4 \pi)^2} + \beta_1 \; {g^5/(4 \pi)^4} } \; \Biggr] 
C_{LL}(m_W) \nonumber\\
=&& {\rm exp}\; \Biggl[\int^{\alpha_S (m_W)}_{\alpha_S (\mu)} \; d\alpha_S ' \; {1 \over 2}
\; \biggl{\lbrace} {\gamma^{(0)} \over {\beta_0} \; \alpha_S ' + 
\beta_1 /(4\,\pi)^2\; {\alpha_S '}^2 }
+ { \gamma^{(1)} \over  \beta_0 \; (4 \; \pi)^2 + \beta_1 \; \alpha_S ' } 
\biggr{\rbrace} \Biggr] 
C_{LL} (m_W)\;,\nonumber\\ 
\end{eqnarray}
where we have used the Eqs. (\ref{beta}) and (\ref{anomdim}). After 
elementary integration, we obtain
\begin{equation}
C_{LL} (\mu) = \overline\eta_{LO} \; 
     \biggl[ 1+ { \alpha_S(m_W) - \alpha_S(\mu) \over
     4 \; \pi } \; \biggl(  {\gamma^{(1)} \over 2 \; \beta_0} -  
     { \gamma^{(0)} \over 2\; \beta_0^2 } \; \beta_1 \biggr) \biggr] \;  
     C_{LL} (m_W) \;,
\end{equation}
where $\overline\eta_{LO}$ is the well known LO-scaling factor given in Eq. 
(\ref{eta0}).

In conclusion we find the following expression for 
$H_{\rm eff}$ at NLO :
\begin{eqnarray}
H_{\rm eff} &=& \frac14{ G_F^2 \over \pi^2} \; m_W^2 \; 
( V_{td} \; V^\ast_{tb} )^2 \; 
\eta_2 (x_W, x_H) \; S_{\rm 2HDM}(x_W, x_H) \tilde{\cal O}_{LL} 
\end{eqnarray}
with
\begin{eqnarray}
\eta_2 (x_W, x_H) &=& {\alpha_S (m_W)}^{\gamma^{(0)}/(2 \;\beta_0)} \;
\Biggl[ 1 + {\alpha_S(m_W) \over 4 \; \pi } \; 
\biggl( {D_{\rm 2HDM}(x_W, x_H) \over S_{\rm 2HDM}(x_W, x_H) } 
+Z\biggr) \Biggr]\;,\\
\tilde{\cal O}_{LL} &=& {\alpha_S(\mu)}^{-\gamma^{(0)}/(2 \;\beta_0)} 
\; \Biggl[ 1 + {\alpha_S(\mu) \over 4 \; \pi } \; 
Z \Biggr] \hat{{\cal O}}_{LL}\;,  
\end{eqnarray}
where
\begin{eqnarray}
Z &=& { \gamma^{(1)} \over 2 \; \beta_0} - \frac{ \gamma^{(0)}}
      {2 \; \beta^2_0}\;\beta_1 \;.
\end{eqnarray}

Neither the factor $\eta_2$, nor the matrix element of $\tilde{\cal O}_{LL}$
are dependent on the low energy scale $\mu\approx m_B$, up to 
${\cal O}(\alpha_s^2)$.

\section{Results}

We want to consider now the mass--splitting $\Delta m_B$ between the 
electroweak mass eigenstates $B_H$ and $B_L$. 
The mass--splitting $\Delta m_B$ is directly given by
the $B^0 \overline {B^0}$--mixing amplitude
\begin{eqnarray}
\nonumber
 \Delta m_B&=&\frac1{m_B}|\langle B^0|H_{\rm eff}|\overline {B^0}\rangle|\\
           &=&\frac{G_F^2}{6\,\pi^2} m_W^2(V_{td}\,V_{tb}^\ast)^2
              S_{\rm 2HDM}(x_W,x_H)\,\eta_2(x_W,x_H)\,B_B\,f_B^2\,m_B\;.
\label{deltam1}
\end{eqnarray}

We obtained the result in Eq. (\ref{deltam1}) by using Eq. (\ref{operator}).
$B_B$ is the renormalization prescription independent B--parameter. It
is defined by \cite{BJW90}

\begin{equation}
B_B=B_B(\mu)\,\alpha_S(\mu)^{-6/23}
    \left[1-\frac{\alpha_S(\mu)}{4\,\pi}\,
         \left(\frac{\gamma^{(1)}}{2\,\beta_0}-\frac{\gamma^{(0)}}
              {2\,\beta^2_0}\,\beta_1   
         \right)
    \right]
\;.
\end{equation}

The term proportional to $\alpha_S$ results from NLO--corrections. We use 
$B_B=1.31\pm0.03$ \cite{BUR96} in our numerical calculation.

The current experimental mean value of the mass-splitting is
given by $\Delta m_B=(3.05\pm0.12)\,\cdot 10^{-13}$ GeV ~\cite{SCH}.
We use the top--quark mass $m_t^{\rm pole}=175\pm 6$ GeV ~\cite{BURAS} 
which corresponds to $m_t(m_W)=176$ GeV.
Furthermore, we employ $m_W=80.33$ GeV, $V_{tb}=0.9991$ ~\cite{PDG} in all 
plots. The meson
decay constant is set to $175\pm25$ MeV ~\cite{BUR96}.

\begin{figure}[h]
\centerline{\mbox{\epsfxsize=12cm\epsffile{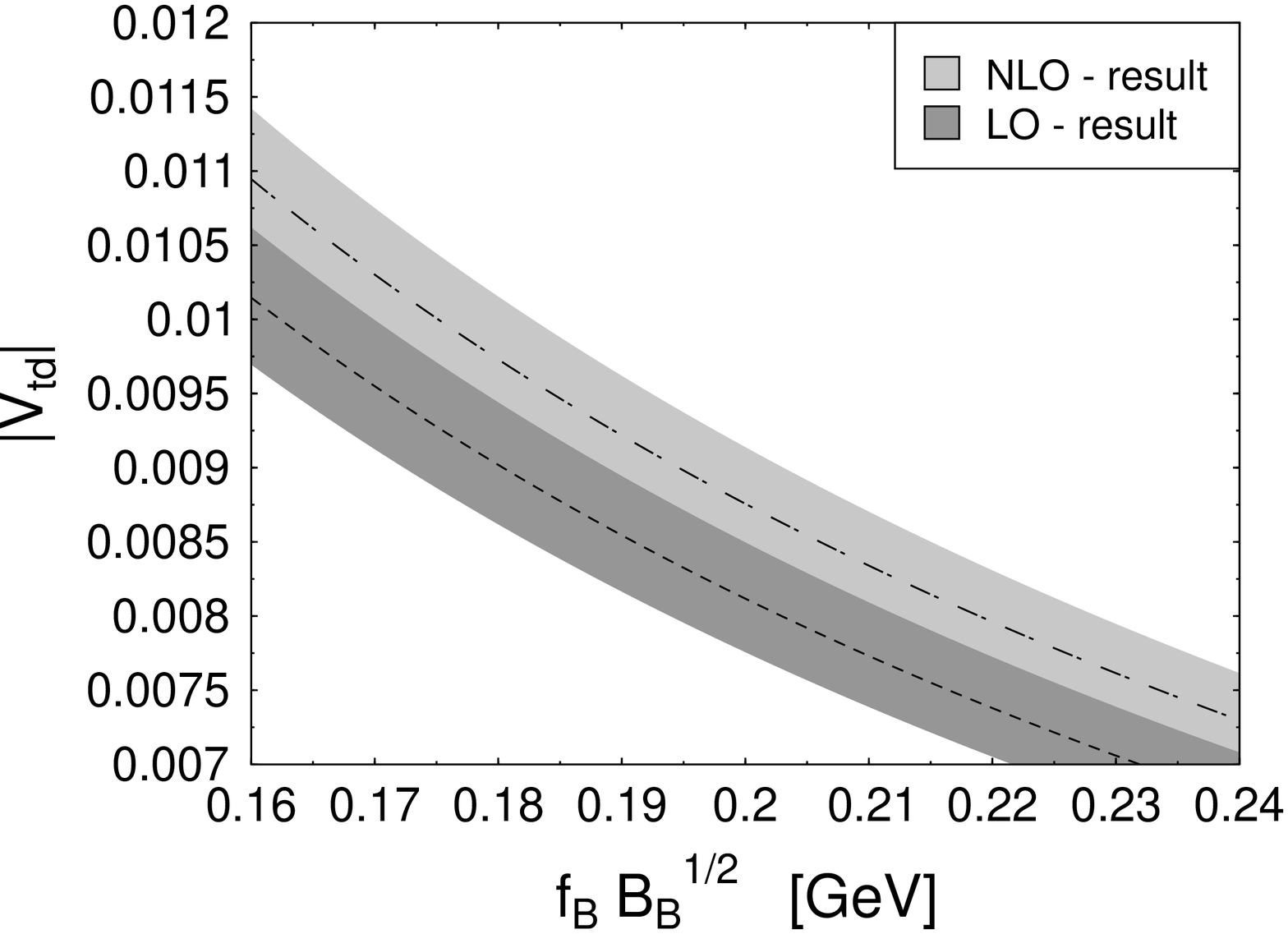}}}
\caption{{\footnotesize
The  CKM--element $V_{td}$ is plotted versus the B meson--decay constant times
the square root of the B--parameter. The shaded areas represent the 
allowed values when errors of $\Delta m_B$ and $m_t$ are included.
There is a small overlap between the LO- and the NLO region shown in the
picture.\label{bild1}}}
\centerline{\mbox{\epsfxsize=12cm\epsffile{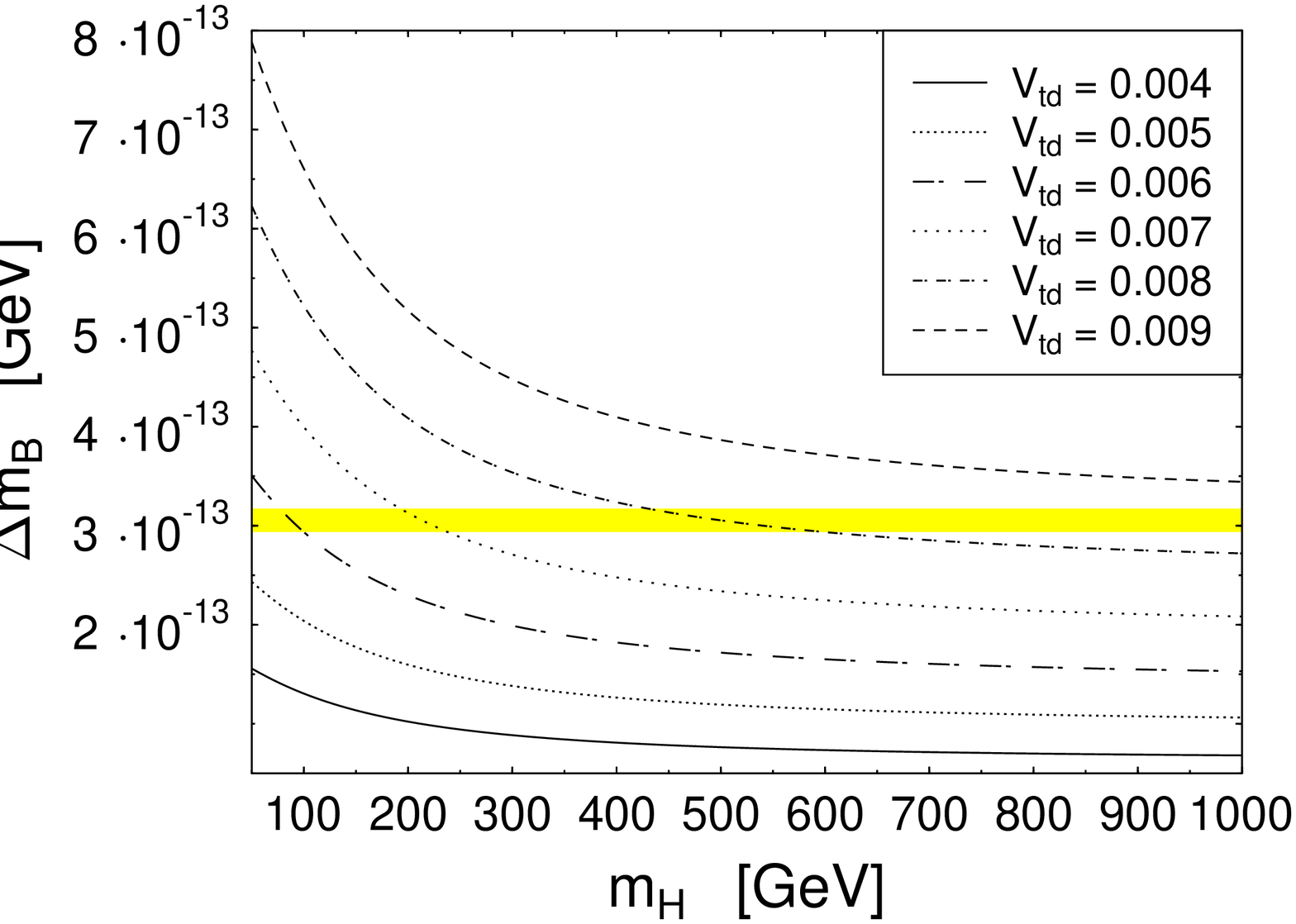}}}
\caption{{\footnotesize
The mass splitting within the 2HDM including NLO-corrections for different
values of $V_{td}$ and for  
$\tan\beta=1$.
The top-quark mass is set to $m_t^{\rm pole}=175$ GeV. The shaded strip is the 
experimentally allowed region. The factor $f_B B_B^{1/2}$ is fixed to 0.2 GeV 
in this figure. The limit $m_H \rightarrow \infty$ yields the 
Standard Model results. \label{bild3}}}
\end{figure}

From Fig. \ref{bild1} we can deduce that the difference between the 
LO--calculation
and the corresponding NLO--calculation for $V_{td}$ is approximately $7\;\%$,
when $m_t^{\rm pole}$ is used in the LO expression. Hence, the 
NLO--calculation enhances the result for the mass--splitting up to $15\;\%$.
This can also be concluded by a direct comparison of 
$\eta_{LO}=\alpha_S(M_W)^{6/23}$
and $\eta_2$ within the SM. We have found $\eta_2=0.4942$ and 
$\eta_{LO}=0.5751$.

If we consider an extended Higgs--sector within the framework of the 2HDM
we have to decrease the related CKM--elements, to get an overlap between
the allowed range for $\Delta m_B$ and the Higgs--mass. 
We recognize for example in Fig. \ref{bild3} that we can not find a physical 
Higgs--boson
with a mass smaller than 1 TeV for $V_{td}=0.0086$, assuming that the
ratio of vacuum expectation values $\tan\beta=1$ and $f_B\,\sqrt{B_B}=0.2$ 
GeV. 
If we decrease $\tan\beta$ we have to decrease $V_{td}$ even more. 
Fig. \ref{bild1} is in full agreement with Fig. \ref{bild3} in the
SM limit $m_H \rightarrow \infty$. 
If we assume $V_{td} = 0.009$ and if we furthermore take into account the
errors for $f_B\,\sqrt{B_B}$ and $m_t$ the 2HDM can not be
distinguished from the SM for Higgs-boson masses larger than 1 TeV.

\begin{figure}[h]
\centerline{\mbox{\epsfxsize=12cm\epsffile{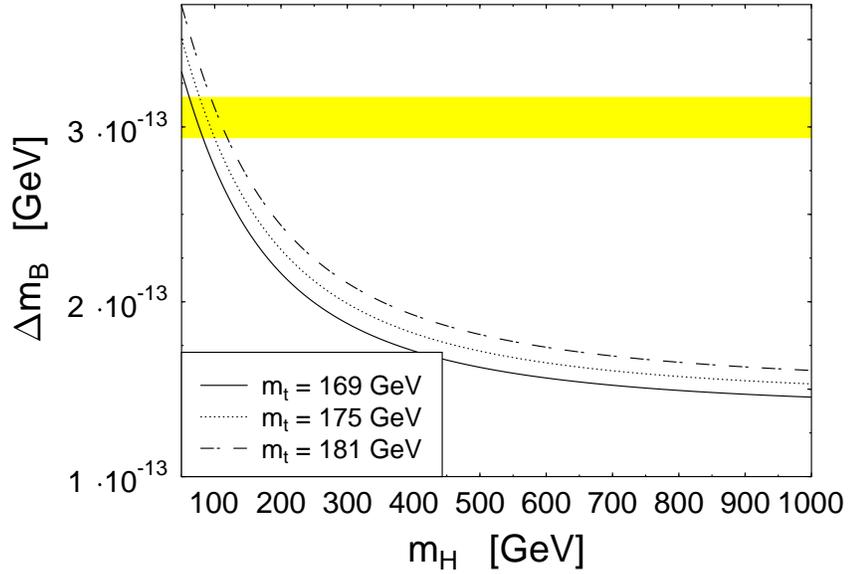}}}
\caption{{\footnotesize
The mass splitting within the 2HDM for different
values of the top-quark pole mass. In this figure, $V_{td}=0.009$, 
$\tan\beta=1$, $f_B B_B^{1/2}=0.2$ GeV.
\label{bild5}}}
\end{figure}

In Fig. \ref{bild5}, it is indicated that we have a relatively sensitive
relation between the top--quark mass and the possible Higgs--boson mass.

\begin{figure}[h]
\centerline{\mbox{\epsfxsize=12cm\epsffile{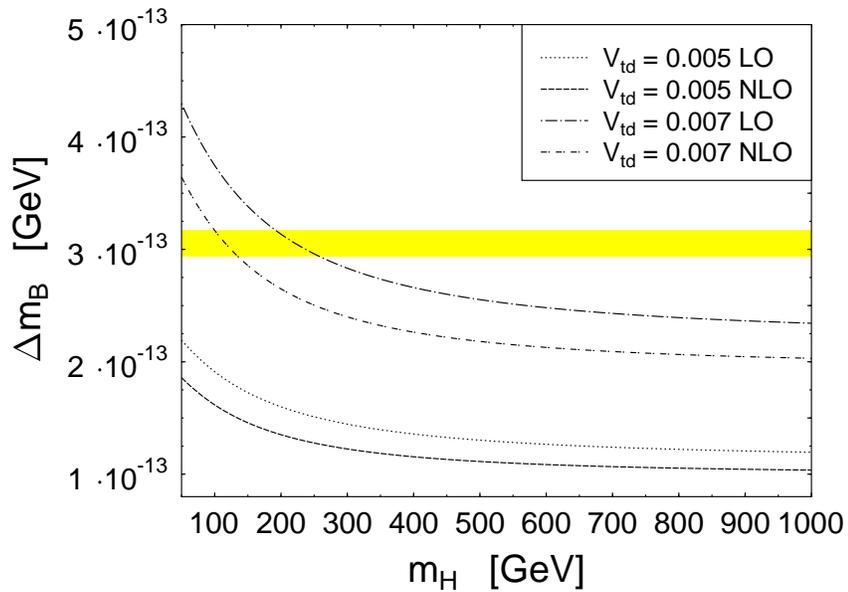}}}
\caption{{\footnotesize
A comparison between LO calculations and NLO-calculations
within the framework of the 2HDM. 
Here, $\tan\beta=1.25$, $f_B B_B^{1/2}=0.2$ GeV and $m_t^{\rm pole}=175$ GeV.
\label{bild6}}}
\end{figure}

The comparison of NLO- and LO-calculation is shown in Fig. \ref{bild6}.
We have plotted the mass-splitting over the Higgs-mass for two typical
values of $V_{td}$ and a ratio of the vacuum expectation values 
$\tan\beta=1.25$. The 
difference between NLO- and LO-calculation approximately amounts to 
$18\;\%$.

\section{Conclusions}

We have calculated the $B^0 \overline{B^0}$--mixing within 
Next--to--Leading--Order with the inclusion of two charged Higgs--bosons.
The NLO calculation leads to an effect of approximately $7\;\%$ 
for $V_{td}$ in comparison to the LO
calculation within the SM. This is indicated in Fig. \ref{bild1}.

We have verified that the scheme developed by Buras, Jamin, Weisz 
~\cite{{BJW90},{BUR90a}} for a proper separation of long
and short distance effects in QCD remains valid in our 
2HDM calculation. This 
can be considered as a good cross--check of the calculations presented here.
The inclusion of NLO-corrections lowers the mass-splitting in the
SM as well as in the 2HDM. We find, for example, a difference between
the pure LO- and the NLO-calculations of approximately $18\;\%$ 
for $m_H=200$ GeV and $17.5\;\%$ for $m_H=400$ GeV. Hence the NLO-contributions
plays an important role in the 2HDM as they correct the LO result
by about $18\;\%$.
 
The inclusion of the two Higgs--bosons leads to an increase of the
calculated amplitude, which depends on the mass
and the vacuum expectation values of two Higgs doublets.
In all practical calculations, the lower bound of the Higgs--boson mass 
was assumed to be 100 GeV and
the upper bound was 1 TeV.

To obtain a mass--splitting within the experimental allowed region, it is
necessary to decrease the CKM--matrix elements 
for small values of $\tan\beta$ and $m_H$. Our calculations
are not valid for higher values of $\tan\beta$ (region near 
$\tan\beta=40$), because it is then necessary to introduce new operators.
  
In the limit of very heavy Higgs--bosons we verify the well--known
results for the Standard Model.

\section*{Acknowledgements}

We are grateful for fruitful discussions with K. Schubert, B. Spaan, 
R. Waldi, Th. Mannel and U. Nierste.
J. U. thanks M. Misiak for valuable
discussions and many useful advices. 
We would like to thank S. Zschocke and Ch. Bobeth for critically 
reading of the manuscript. We acknowledge support by DFG, GSI (Darmstadt) 
and BMBF.


\mathindent0pt

\appendix

\section*{Appendix A}
\subsection*{A.1 Special functions}
After carrying out the second integration, the dilogarithm or Spence
function appears in our results. It is defined by
\begin{equation}
{\rm Li}_2\,(x)=-\int_0^x\,dt\,\frac{\ln(1-t)}t=\sum_{n=1}^\infty
\frac{x^n}{n^2}\,;\quad |x|<1\;\;.
\end{equation}
The following useful relations hold
\begin{equation}
{\rm Li}_2\,(1-x)+{\rm Li}_2\,(1-1/x)=-\frac12\ln^2(x)\;\;,
\end{equation}
\begin{equation}
{\rm Li}_2\,(x)+{\rm Li}_2\,(1-x)=\frac{\pi^2}6-\ln(x)\,\ln(1-x)\;\;.
\end{equation}

\subsection*{A.2 Inami--Lim functions and NLO in the SM}

We list now the Inami--Lim functions and the well--known 
functions $L^{(i,SM)}(x_W)$.
\begin{eqnarray}
S_{WW}(x_W) &=& x_W\;\left(\frac14+\frac9{4(1-x_W)}-\frac3{2(1-x_W)^2}-
                \frac{3x_W^2\ln (x_W)}{2\,(1-x_W)^3}\right)\;,\\
S_{HH}(x_H) &=& \frac{1}{\tan^4(\beta)}\,\frac{x_H\;x_W}4\left(\frac{1+x_H}{(1-x_H)^2}+
                \frac{2x_H\ln(x_H)}{(1-x_H)^3}\right)\;,\\
\nonumber
2\;S_{WH}(x_W,x_H)&=&\frac{1}{\tan^2(\beta)}\,\frac{x_H\;x_W}4\left(\frac{(2 x_W -8 x_H)
                  \ln(x_H)}{(1-x_H)^2
                  (x_H-x_W)}+\frac{6x_W\ln(x_W)}{(1-x_W)^2(x_H-x_W)}\right.\\
&&\qquad\qquad\qquad\left.-\frac{8-2x_W}{(1-x_H)(1-x_W)}\right)\quad .
\end{eqnarray}
The function $S_{WH}(x_W,x_H)$ is obtained by including the charged
would--be Goldstone bosons or in a more direct way, by choosing 
the unitary gauge
for the $W$--boson right from the beginning.

The functions $L^{(i,SM)}$ can be decomposed as  
\begin{equation}
L^{(i,SM)}(x_W) = WW^{(i)}(x_W)\,+\,2\, W\Phi^{(i)}(x_W)\,+\,\Phi\Phi^{(i)}(x_W)
\end{equation}
where
\begin{eqnarray}
WW^{(1)}(x_W)&=& WW^{(1)}_{tt} (x_W) -2 \; WW^{(1)}_{tu}(x_W) 
+ WW^{(1)}_{uu} (x_W) \;\;,
\end{eqnarray}

\begin{eqnarray}
WW^{(1)}_{tt} (x_W) & = & { (4 \; x_W + 38 \; x_W^2 +6 \; x_W^3) \; \ln (x_W) 
	\over {(x_W-1)}^4 } 
	+ { (12 \; x_W +  48 \; x_W^2 + 12 \; x_W^3 ) \;{\rm Li}_2 ( 1- { 1 / x_W}) 
	\over  {(x_W-1)}^4 } \nonumber \\[2mm]
& & + { (24 \; x_W + 48 \; x_W^2) \; {\rm Li}_2 ( 1- x_W )
	 \over {(x_W-1)}^4 }
	- { { 3+ 28 \; x_W + 17 \; x_W^2}
	 \over {(x_W-1)}^3 }\;\;,
\end{eqnarray}

\begin{eqnarray}
2 \, WW^{(1)}_{tu} (x_W) & = &  \frac{ 2\,(3\,+\,13\,x_W)} {(x_W-1)^2 } 
	-\frac{ 2\,x_W\,(5 \,+\, 11\,x_W) \; \ln (x_W)}
	      {(x_W-1)^3 } \nonumber \\ [2mm]
& & - \frac{12\,x_W\,(1\,+\,3 \, x_W) \; {\rm Li}_2 ( 1-  1/x_W  )} 	
	   {(x_W-1)^3 }
    - \frac{ 24\,x_W\,(1+x_W){\rm Li}_2 ( 1-  x_W  )}
	   {(x_W-1)^3}\;,
\end{eqnarray}

\begin{eqnarray}
WW^{(1)}_{uu}(x_W) & =& 3 \;\;,
\end{eqnarray}

\begin{eqnarray}
\Phi\Phi^{(1)}(x_W)&=&-\frac{x_W^2\,(7+52\,x_W-11\,x_W^2)}
              {4\,(x_W-1)^3}+\frac{3\,x_W^3\,(4+5\,x_W-x_W^2)\ln(x_W)}
              {2\,(x_W-1)^4}\nonumber\\[2mm]
& &+\frac{3\,x_W^3\,(3+4\,x_W-x_W^2){\rm Li}_2(1-1/x_W)}
   {(x_W-1)^4}+\frac{18\,x_W^3\,{\rm Li}_2(1-x_W)}{(x_W-1)^4}\;\;,
\end{eqnarray}

\begin{eqnarray}
\nonumber
2\;W\Phi^{(1)}(x_W)&=&\frac{4\,x_W^2\,(11+13\,x_W)}
              {(x_W-1)^3}+\frac{2\,x_W^2\,(5\,+\,x_W)(1-9\,x_W)\ln(x_W)}
              {(x_W-1)^4}\\[2mm]
& &-\frac{24\,x_W^2\,(1+4\,x_W+x_W^2){\rm Li}_2(1-1/x_W)}
   {(x_W-1)^4}-
   \frac{48\,x_W^2\,(1+2\,x_W){\rm Li}_2(1-x_W)}{(x_W-1)^4}\;\;,\nonumber\\ 
\end{eqnarray}

\begin{eqnarray}
WW^{(8)}(x_W)&=& WW^{(8)}_{tt} (x_W) -2 \; WW^{(8)}_{tu}(x_W) + 
WW^{(8)}_{uu} (x_W)\;\;, 
\end{eqnarray}

\begin{eqnarray}
WW^{(8)}_{tt} (x_W) & = & { 2\; x_W (4 \;-\;3 \; x_W)\;
         \ln (x_W)\over {(x_W-1)}^3 } 
	- { (12 \; x_W -12 \; x_W^2 - 8 \; x_W^3 ) \;{\rm Li}_2 
         ( 1- { 1 / x_W}) 
	\over  {(x_W-1)}^4 } \nonumber \\[2mm]
& & + { (8- 12 \; x_W + 12 \; x_W^2) \;{\rm Li}_2 ( 1- \; x_W ) \over  
        {(x_W-1)}^4}
	- { {(23 \;- \; x_W)} \over {(x_W-1)}^2}\;\;
\end{eqnarray}

\begin{eqnarray}
2 \; WW^{(8)}_{tu} (x_W) & = &  { 2\,(2\,-\,x_W)
        \; \pi^2 \over 3\,x_W} 
	- { (8\,-\,5 \; x_W ) \;\ln(x_W) 
	 \over {(x_W-1)}^2 } \nonumber \\ [2mm]
& & -  { (6 \; x_W\,+\,4 \; x_W^2  ) \;{\rm Li}_2 ( 1- {1 / x_W } )
	 \over x_W\,  {(x_W-1)}^2}
	+   { (8\,+\,12\;x_W\,-\,6 \; x_W^2) \;{\rm Li}_2 ( 1-  x_W  ) 
	\over x_W\, {(x_W-1)}^2 } \nonumber \\[2mm]
& &	-  { {15} \over {(x_W-1)}}\;\;,
\end{eqnarray}

\begin{eqnarray}
WW^{(8)}_{uu} (x_W)  &=& -23 + { 4 \over 3 }\; \pi^2\;\;,
\end{eqnarray}

\begin{eqnarray}
\nonumber
\Phi\Phi^{(8)}(x_W)&=&-\frac{11\,x_W^2\,(1+x_W)}
              {4\,(x_W-1)^2}+\frac{x_W^3\,(4\,-3\,x_W)\ln(x_W)}
              {2\,(x_W-1)^3}\\
& &+\frac{x_W^3\,(3-3\,x_W+2\,x_W^2){\rm Li}_2(1-1/x_W)}
   {(x_W-1)^4}+
   \frac{x_W^2(2+3\,x_W-3\,x_W^2){\rm Li}_2(1-x_W)}{(x_W-1)^4}\;\;,\nonumber\\
\end{eqnarray}

\begin{eqnarray}
\nonumber
2\;W\Phi^{(8)}(x_W)&=&\frac{30\,x_W^2}{(x_W-1)^2}
              +\frac{12\,x_W^3\ln(x_W)}
              {(x_W-1)^3}-\frac{12\,x_W^4\,{\rm Li}_2(1-1/x_W)}
              {(x_W-1)^4}\\
& & -\frac{12\,x_W^2\,(2\;-\,x_W^2){\rm Li}_2(1-x_W)}{(x_W-1)^4}\;\;. 
\end{eqnarray}
\subsection*{A.3 NLO in the extended Higgs sector}

The $L^{(i,H)}$ can be decomposed in the following way
\begin{eqnarray}\nonumber
L^{(i,H)}(x_W,x_H) = \frac{2}{\tan^2(\beta)}\,WH^{(i)}(x_W,x_H)\,+
\frac{2}{\tan^2(\beta)}\,
\Phi H^{(i)}(x_W,x_H)+\,\frac{1}{\tan^4(\beta)}\,HH^{(i)}(x_H)\;.
\\
\end{eqnarray}
We list now the newly obtained results for the explicit QCD corrections
in the extended Higgs sector:

\begin{eqnarray}
HH^{(1)}(x_i)&=&\frac{x_W}{x_H}\Phi\Phi^{(1)}(x_H) + 6 \;(\ln (x_H)
                - \ln (x_W))\; \sum_{i=H,W} x_i \frac{\partial}{\partial x_i}
                 \; S_{HH} (x_i)\;\;,
\end{eqnarray}

\begin{eqnarray}\nonumber
2\;WH^{(1)}(x_i)
&=&x_W\left(\frac{2\,x_H^2(13+3\,x_H)\ln(x_H)}
              {(x_H-1)^3(x_H-x_W)}
   -\frac{2\,x_H\,(9+7\,x_H+7\,x_W-23\,x_W\,x_H)}
              {(x_W-1)^2(x_H-1)^2}\right.\\
\nonumber
& &-\frac{2\,x_H^2(18\,-\,6\,x_H\,-44\,x_W+13\,x_H\,x_W+
   9\,x_H\,x_W^2)\ln(x_W)}
   {(x_H-1)^2(x_W-1)^3(x_H-x_W)}\\
\nonumber
& &+\frac{2\,x_H\,x_W(5-
   27\,x_W+6\,x_W^2+6\,x_H\,x_W^2)\ln(x_W)}
              {(x_H-1)^2(x_W-1)^3(x_H-x_W)}-\frac{24\,x_H^2\,\ln(x_H)\,
   \ln(x_W)}{(x_H-1)^3(x_H-x_W)}\\
\nonumber
& &+\frac{24\,x_H^2\,\rm{ Li}_2(1-1/x_H)}{(x_H-1)^2(x_H-x_W)}-
   \frac{24\,x_H\,x_W\,(1+x_W)\,\rm{ Li}_2(1-1/x_W)}{(x_W-1)^3(x_H-x_W)}\\
& &\left. -\frac{48\,x_W\,x_H\,\rm{ Li}_2(1-x_W)}{(x_W-1)^3(x_H-x_W)}\right)
\;\;, 
\end{eqnarray}

\begin{eqnarray}\nonumber
2\;\Phi H^{(1)}(x_i)
&=&x_W^2\left(\frac{x_H\,(31-15\,x_H\,-\,15\,x_W-x_H\,x_W)}
                   {2\,(x_H\,-\,1)^2\,(x_W\,-\,1)^2}
   -\frac{x_H\,(7+21\,x_H\,-\,12\,x_H^2)\,\ln(x_H)}{2\,(x_H\,-\,1)^3\,
    (x_H\,-\,x_W)}\right.\\
\nonumber
& &+\frac{x_H\,(7\,-\,9\,x_W\,+\,36\,x_W^2\,-\,18\,x_W^3)\,
    \ln(x_W)}{2(x_H-1)^2(x_H-x_W)(x_W-1)^3}\\
\nonumber
& &+\frac{x_H^2\,(8\,-\,36\,x_W\,+\,9\,x_W^2\,+\,3\;x_W^3)
    \ln(x_W)}{(x_H-1)^2(x_H-x_W)(x_W-1)^3}\\
\nonumber
& &-\frac{x_H^3\,
    (11\,-\,45\,
    x_W\,+\,18\,x_W^2)\,\ln(x_W)}{2(x_H-1)^2(x_H-x_W)(x_W-1)^3}
    +\frac{6\,x_H\,\ln(x_H)\,\ln(x_W)}{(x_H-1)^3(x_H-x_W)}\\
\nonumber
& &-\frac{6\,x_H(1+x_H-x_H^2){\rm Li}_2(1-1/x_H)}
    {(x_H-1)^2(x_H-x_W)}\\
& &\left.+\frac{6\,x_H\,(1\,+\,2\,x_W^2\,-\,x_W^3){\rm Li}_2(1-1/x_W)}
         {(x_H\,-\,x_W)(x_W\,-\,1)^3}
   +\frac{12\,x_H\,{\rm Li}_2(1-x_W)}{(x_H\,-\,x_W)(x_W\,-\,1)^3}\right)
   \;\;,
\end{eqnarray}     

\begin{eqnarray}
HH^{(8)}(x_i)&=&\frac{x_W}{x_H}\Phi\Phi^{(8)}(x_H) + 6 \;( \ln(x_H) -
                \ln(x_W)) \;S_{HH} (x_i)\;\;,
\end{eqnarray}

\begin{eqnarray}
2\;WH^{(8)} (x_i) & = &
\nonumber
x_W \left(
\frac{24 x_H x_W {\rm Li_2}(1 - x_W)}{(x_H - x_W) (x_W-1)^2} \right.\\
& &\nonumber
+\frac{6 x_H^2 (5 x_W - x_H + 3 x_W^2 x_H) {\rm Li_2} (1-1/x_W)}
{(x_H-1)^2 (x_H - x_W) (x_W-1)^2 x_W} \\
& &\nonumber
+\frac{6 x_H (2 x_W^2 - 10 x_H x_W + x_H x_W^2) {\rm Li_2} (1-1/x_W)}
{(x_H-1)^2 (x_H - x_W) (x_W-1)^2} \\
& &\nonumber
+\frac{6 x_H^2 (5 x_W - x_H -8 x_W^2+ 2 x_H x_W^2) {\rm Li_2} (1-x_H)}
{(x_H-1)^2 (x_H - x_W) (x_W-1)^2 x_W} \\
& &\nonumber
+\frac{6 (x_W^2 - x_H x_W + 2 x_H^2 x_W^2) {\rm Li_2}(1-x_H)} 
{(x_H-1)^2 (x_H - x_W) (x_W-1)^2} \\
& &\nonumber
+\frac{6 x_H^2 (-x_H+5 x_W) {\rm Li_2}(1-1/x_H)}
{(x_H-1)^2 (x_H - x_W) x_W} \\
& &\nonumber
-\frac{6 x_H^2 (5 x_W - x_H -8 x_W^2+ 2 x_H x_W^2) {\rm Li_2} (1-x_H/x_W)}
{(x_H-1)^2 (x_H - x_W) (x_W-1)^2 x_W} \\
& &\nonumber
-\frac{6 (x_W^2 - x_H x_W + 2 x_H^2 x_W^2) {\rm Li_2}(1-x_H/x_W)} 
{(x_H-1)^2 (x_H - x_W) (x_W-1)^2} \\
& &\nonumber
-\frac{6 x_H (1-x_H - \ln (x_H))}{(x_H-1)^2 (x_W-1)}
+\frac{6 x_H (2 x_W-1) \ln(x_W)}
{(x_H-1) (x_W-1)^2} \\
& &\nonumber
+\frac{6 x_H^2 (5 x_W - x_H - 8 x_W^2) \ln(x_H) \ln(x_W)}
{(x_H-1)^2 (x_H - x_W) (x_W-1)^2 x_W} \\
& &\left.
+\frac{12 x_H^2 (x_H x_W +x_W^2) \ln(x_H) \ln(x_W)}
{(x_H-1)^2 (x_H - x_W) (x_W-1)^2}\right)\;,
\end{eqnarray}

\begin{eqnarray}
\nonumber
2\;\Phi H^{(8)}(x_i)
&=&x_W^2\left(\frac{2\;x_H\;+\;2\;x_W\;-\;11\;x_H\;x_W}
                   {2\;( x_H - 1)\;( x_W - 1)\;x_W}\right.\\
\nonumber
& &-\frac{(2\;x_H^2\;-\;7\;x_H\;x_W\;+\;2\;x_H^2\;x_W\;+\;2\;
          x_W^2\;+\;x_H\;x_W^2)
          \ln (x_H)}{2\;( x_H - 1)^2\;(x_H - x_W)\;( x_W - 1)\;x_W}\\
\nonumber
& &-\frac{x_H\;(7\;-\;7\;x_H\;+\;4\;x_W\;-\;6\;x_W^2)\;\ln (x_W)}
          {2\;( x_H - 1)\;(x_H - x_W)\;( x_W - 1)^2}\\
\nonumber
& &+\frac{(x_H^2\;+\;x_W^2\;-\;3\;x_H^2\;x_W^2)\;\ln (x_W)}
          {( x_H - 1)\;(x_H - x_W)\;( x_W - 1)^2\;x_W}\\
\nonumber
& &-\frac{x_H^2\;(4\;-\;6\;x_W\;+\;3\;x_H\;x_W)\; \ln (x_H)\; \ln (x_W)}
          {( x_H - 1)^2\;(x_H - x_W)\;( x_W - 1)^2\;x_W}\\
\nonumber
& &+\frac{x_H\;(x_H^2\;-\;3\;x_W^2\;+\;6\;x_W^3\;-\;3\;x_W^4)\; \ln (x_H)\; \ln (x_W)}
          {( x_H - 1)^2\;(x_H - x_W)\;( x_W - 1)^2\;x_W^2}\\
\nonumber
& &-\frac{x_H\;(3\;x_W^2\;+\;2\;x_H\;x_W\;(2\;+\;x_W)\;-\;x_H^2\;(1\;+\;2\;
          x_W))\;{\rm Li}_2(1-1/x_H)}{( x_H - 1)^2\;(x_H - x_W)\;x_W^2}\\
\nonumber
& &-\frac{(4\;x_H\;x_W\;-\;6\;x_H^2\;x_W\;+\;3\;x_H^2\;x_W^2\;-\;x_W^2)
          {\rm Li}_2(1-x_H)}{( x_H - 1)^2\;(x_H - x_W)\;( x_W - 1)^2\;x_H}\\
\nonumber
& &-\frac{(4\;x_H^2\;x_W\;-\;6\;x_H^2\;x_W^2\;-\;x_H^3\;+\;3\;x_H^3\;x_W^2)
          {\rm Li}_2(1-x_H)}{( x_H - 1)^2\;(x_H - x_W)\;( x_W - 1)^2\;x_W^2}\\
\nonumber
& &+\frac{2\;x_H^2\;(6\;-\;x_W^2\;-\;3\;x_H\;+\;x_W\;x_H)\;
          {\rm Li}_2(1-1/x_W)}{( x_H - 1)^2\;(x_H - x_W)\;( x_W - 1)^2}\\
\nonumber
& &-\frac{x_H\;(3\;x_W^2\;+\;4\;x_H\;x_W\;-\;x_H^2)\;{\rm Li}_2(1-1/x_W)}
          {( x_H - 1)^2\;(x_H - x_W)\;( x_W - 1)^2\;x_W^2}\\
\nonumber
& &+\frac{(4\,x_H\;x_W\;-\;6\;x_H^2\;x_W\;+\;3\;x_H^2\;x_W^2\;-\;x_W^2)
           {\rm Li}_2(1-x_H/x_W)}{( x_H - 1)^2\;(x_H - x_W)\;( x_W - 1)^2\;
            x_H}\\
\nonumber
& &+\frac{x_H^2(4\;x_W\;-\;6\;x_W^2\;-\;x_H\;+\;3\;x_H\;x_W^2)
           {\rm Li}_2(1-x_H/x_W)}{( x_H - 1)^2\;(x_H - x_W)\;( x_W - 1)^2\;
            x_W^2}\\
& &\left.-\frac{6\;x_H\;{\rm Li}_2(1-x_W)}{(x_H-x_W)\;(x_W-1)^2}\right)\;.
\end{eqnarray}

We have used the capital $W$, $\Phi$ and $H$ to label the W--boson, the
charged would--be Goldstone boson
and the physical Higgs--boson, respectively, 
i.e. $W\Phi$ denotes a diagram with one internal $W$--boson and one internal 
unphysical scalar.

\newpage


\newpic{1}

\newpic{3}

\newpic{4}

\newpic{4}

\newpic{6}

\newpic{9}

\end{document}